\documentclass[9pt,twocolumn,twoside]{pnas-new}
\pdfoutput=1
\usepackage{url}
\usepackage{pdfpages}
\usepackage{import}

\templatetype{pnasresearcharticle} 

\title{The Chaperone Effect in Scientific Publishing}

\author[a]{Vedran Sekara}
\author[b,c]{Pierre Deville}
\author[d]{Sebastian Ahnert}
\author[c,e,f,g,*]{Albert-L\'{a}szl\'{o} Barab\'{a}si}
\author[e,c,h,i,1,*]{Roberta Sinatra} 
\author[a,j,1,*]{Sune Lehmann}

\affil[a]{Department of Applied Mathematics and Computer Science, Technical University of Denmark, Kgs. Lyngby, Denmark}
\affil[b]{Department of Applied Mathematics, Universit\'{e} catholique de Louvain, Belgium}
\affil[c]{Center for Complex Network Research, Northeastern University, Boston, USA}
\affil[d]{Theory of Condensed Matter, Cavendish Laboratory, University of Cambridge, United Kingdom}
\affil[e]{Center for Network Science and Department of Mathematics, Central European University, Budapest, Hungary}
\affil[f]{Center for Cancer Systems Biology, Dana-Farber Cancer Institute, Boston, USA}
\affil[g]{Channing Division of Network Medicine, Brigham and Women’s Hospital, Harvard Medical School, Boston, USA}
\affil[h]{Complexity Science Hub, Vienna, Austria}
\affil[i]{ISI Foundation, Torino, Italy}
\affil[j]{The Niels Bohr Institute, University of Copenhagen, Copenhagen, Denmark}

\leadauthor{Sekara} 

\authorcontributions{S.L. conceived the study.  V.S., R.S. and S.L designed measures and analyses. V.S., P.D. and R.S. collected and curated the data. V.S. conducted the analysis.  S.A. made the analytical calculations to explain the alphabetical null model. R.S., S.L., V.S. and A-L.B. wrote the manuscript. S.L. and R.S lead the research. All authors interpreted and discussed the findings.}
\authordeclaration{The authors have no conflict of interest.}
\equalauthors{\textsuperscript{1}S.L. and R.S. lead the research and contributed equally to this work.}
\correspondingauthor{\textsuperscript{*}To whom correspondence should be addressed: alb@neu.edu, robertasinatra@gmail.com, sunelehmann@gmail.com}


\begin{abstract}
Experience plays a critical role in crafting high impact scientific work. This is particularly evident in top multidisciplinary journals, where a scientist is unlikely to appear as senior author if they have not previously published within the same journal. 
Here, we develop a quantitative understanding of author order by quantifying this ‘Chaperone Effect’, capturing how scientists transition into senior status within a particular publication venue. 
We illustrate that the chaperone effect has different magnitude for journals in different branches of science, being more pronounced in medical and biological sciences and weaker in natural sciences. 
Finally, we show that in the case of high-impact venues, the chaperone effect has significant implications, specifically resulting in a higher average impact relative to papers authored by new PIs. Our findings shed light on the role played by experience in publishing within specific scientific journals, on the paths towards acquiring the necessary experience and expertise, and on the skills required to publish in prestigious venues.
\end{abstract}

\begin{document}

\verticaladjustment{-2pt}

\maketitle
\thispagestyle{firststyle}
\ifthenelse{\boolean{shortarticle}}{\ifthenelse{\boolean{singlecolumn}}{\abscontentformatted}{\abscontent}}{}

\dropcap{S}cience as an institution is highly stratified~\cite{cole1973social}, and anecdotal evidence that scientific high-achievers are often prot\'eg\'es of accomplished mentors, supports the notion that scientific status is passed along through lineages of prominent scientists~\cite{math_genealogy, physics_genealogy, parberry2004sigact}. 
While single-topic studies like the mathematical genealogy project document such bonds between renowned scientists~\cite{malmgren2010role}, there is less quantitative understanding of the role of apprenticeship in scientific publishing and of how scientific excellence is passed along between generations~\cite{chao1992formal,malmgren2010role}.
Here we quantify a key aspect of this `chaperone effect' by considering how inexperienced scientists transition into senior status given multiple publications within the same scientific journal.
We illustrate that the chaperone effect has different magnitude for journals in different branches of science, the effect being more pronounced within medical and biological sciences and weaker for the natural sciences.
For high-impact multidisciplinary journals, a scientist is unlikely to appear as senior author if he or she has not previously published within the same journal.
Our findings shed light on the role played by scientific training to acquire the necessary experience, expertise and skills to publish in venues characterized by a strong chaperone effect. 

\begin{figure*}
\centering
\includegraphics[width=0.99\linewidth]{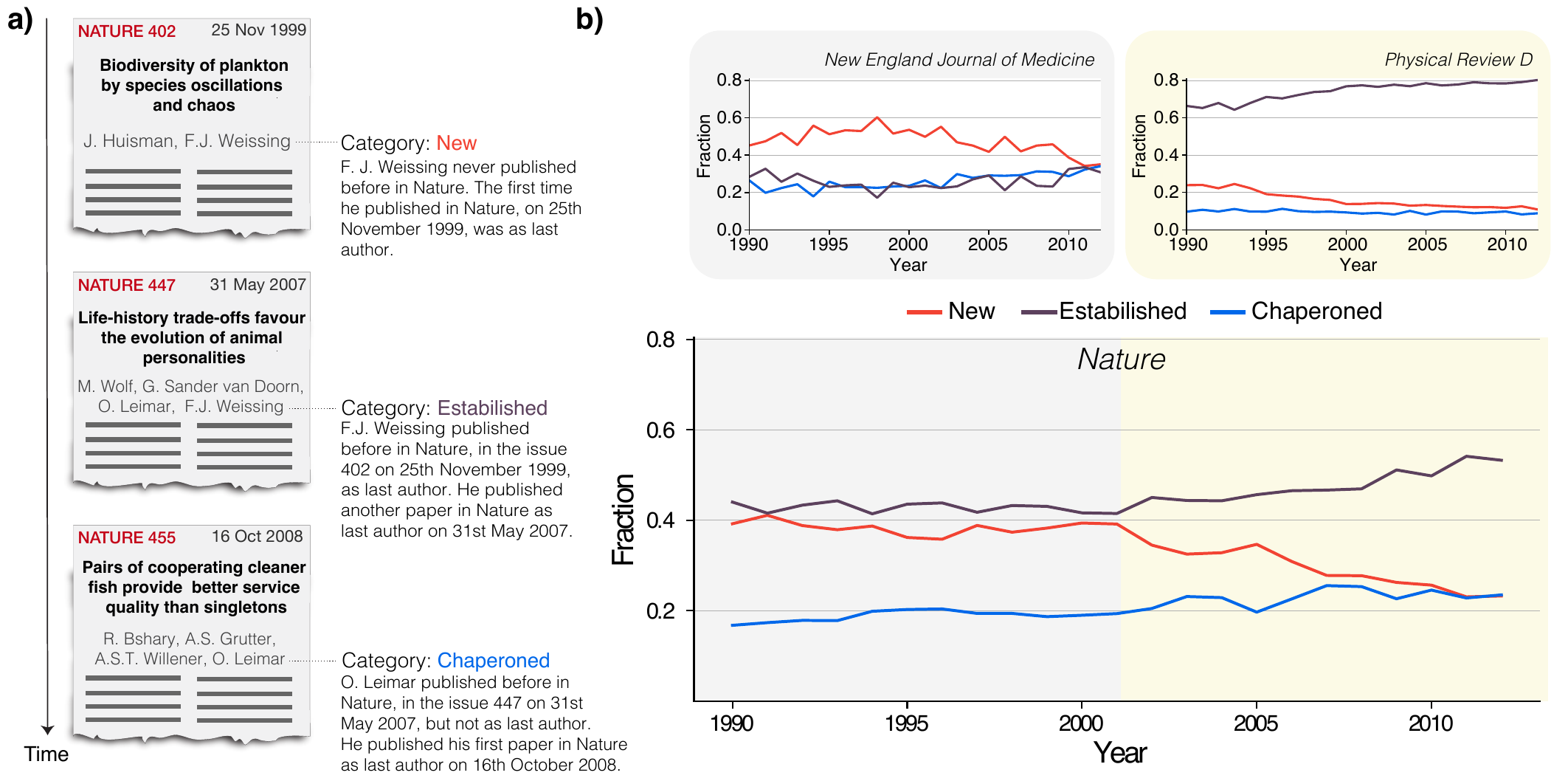}
\caption{\textbf{Probability of being listed as PI in Nature given previous publication history.} 
\textbf{a}, Terminology of authors. The last authors of all papers published each year in Nature are divided into three categories: \textit{new} authors, that have never published in \textit{Nature} before, \textit{chaperoned} authors, that have published in \textit{Nature} before only at junior level, and \textit{established} authors, that have already previously published as last authors.
\textbf{b}, Change in author fractions over time for three journals, displaying different trends over time: while in the \textit{New England Journal of Medicine (NEJM)} the proportion of different PIs tends to be equally balanced over time, in \textit{Physical Review D} this proportion tends to become more unbalanced, with the fraction of established PIs increasing. For author fractions in \textit{PNAS} see SI Appendix fig.~S2.}
\label{fig:last_author_fraction}
\end{figure*}

In general, there are a wealth of indications that young scientists who interact with successful mentors have a higher probability of achieving success later in their careers.
For example, an improbably large fraction of Nobel laureates were trained by other laureates~\cite{cole1973social,zuckerman1967nobel}. 
Beyond the core skill of learning to select relevant scientific questions and providing meaningful answers, an important aspect of career success rests on publishing in prestigious venues.
Here we focus, not on mentorship directly, but on an important facet of the mentorship process: Experience with publishing within a specific journal. 

The order of authors on multi-author scientific articles provides important signals regarding the role of each scientist in a project  \cite{lehmann2006measures, shen2014collective}. 
For example, in biological and increasingly in physical sciences typically, the first author is an early-career scientist who carries out the research, while the last author is a mentor-figure who plays a role in shaping the research, establishing the paper's structure, and corresponding with journal editors~\cite{van2014publication,Austin20062014}.
Middle authors generally play more specialized roles, such as contributing statistical analyses.
This division of labor is often symbiotic; it has recently been shown that junior researchers tend to work on more innovative topics but need mentorship~\cite{packalen2015age,callaway2015young}.
Further, high impact works are often performed by multiple authors whose composition is usually heterogeneous in terms of experience~\cite{wuchty2007increasing,jones2008multi,milojevic2014principles,klug2016understanding}.
In this work, we use author-order to study the role of experience in crafting scientific work~\cite{chao1992formal} by analyzing the dynamics of scientific multi-author publications~\cite{van2014publication,Austin20062014}.
Such sequences provide a `petri dish', unveiling the patterns that increase the rate of acceptance for some authors.
To unravel how the dynamics of these sequences vary across the sciences, we explore the extent to which the principal investigator (PI) of a paper has previously published in the same journal as a junior author. 
Thus, we address a question which is often asked by scientists: `Can you publish in \emph{Nature} if you have never published in \emph{Nature} before!?'.
Note that here we take \textit{Nature} as an example of a journal with high impact factor.
However, our analysis spans multiple journals, as described in Materials and Methods.

We consider $6.1$ million papers published between 1960 and 2012 in $386$ scientific journals; covering the fields of mathematics, physics, chemistry, biology, and medicine (see Materials and Methods for data processing and name disambiguation). 
Included are the top $3$ multidisciplinary journals: \emph{Nature}, \emph{Science}, and \emph{PNAS}. 
In our analysis, we assume that the principal investigator (PI) is listed last in a paper's author list, a common practice in many scientific fields~\cite{waltman2012empirical, van2014publication,Austin20062014}.
Note, however, that our analysis is not affected if the author list of some papers does not mirror seniority roles (see Materials and Methods).
For all papers in each journal, we divide PIs into 3 categories: \textit{new} PIs are those who have not published previously in that specific journal, \textit{chaperoned} PIs are those who have appeared before only as junior (non-last) authors, and \textit{established} PIs are those already previously listed as a last author in the journal (see Fig.~1a). By definition, the last author on any given publication can be classified only in one of these categories. 
For example, F.J.~Weissing's first paper in \textit{Nature} was as last author, so he is labeled as a \textit{new} PI in \textit{Nature} for that year (1999). 
In 2007, Weissing published in \textit{Nature} as last-author again, but because of the previous publication, we categorize him as as an \textit{established} PI in \textit{Nature} in 2007. 
This 2007 \textit{Nature} paper was co-authored by three other scientists, one of them being O.~Leimar. 
A year later, Leimar published a paper as last author in \textit{Nature}, and is therefore marked as a \textit{chaperoned} PI in \textit{Nature} for that year. 
In Figure~1b we show the fraction of \emph{new}, \emph{established}, and \emph{chaperoned} authors over time for three scientific journals. 

\begin{figure*}
\centering
\includegraphics[width=0.99\linewidth]{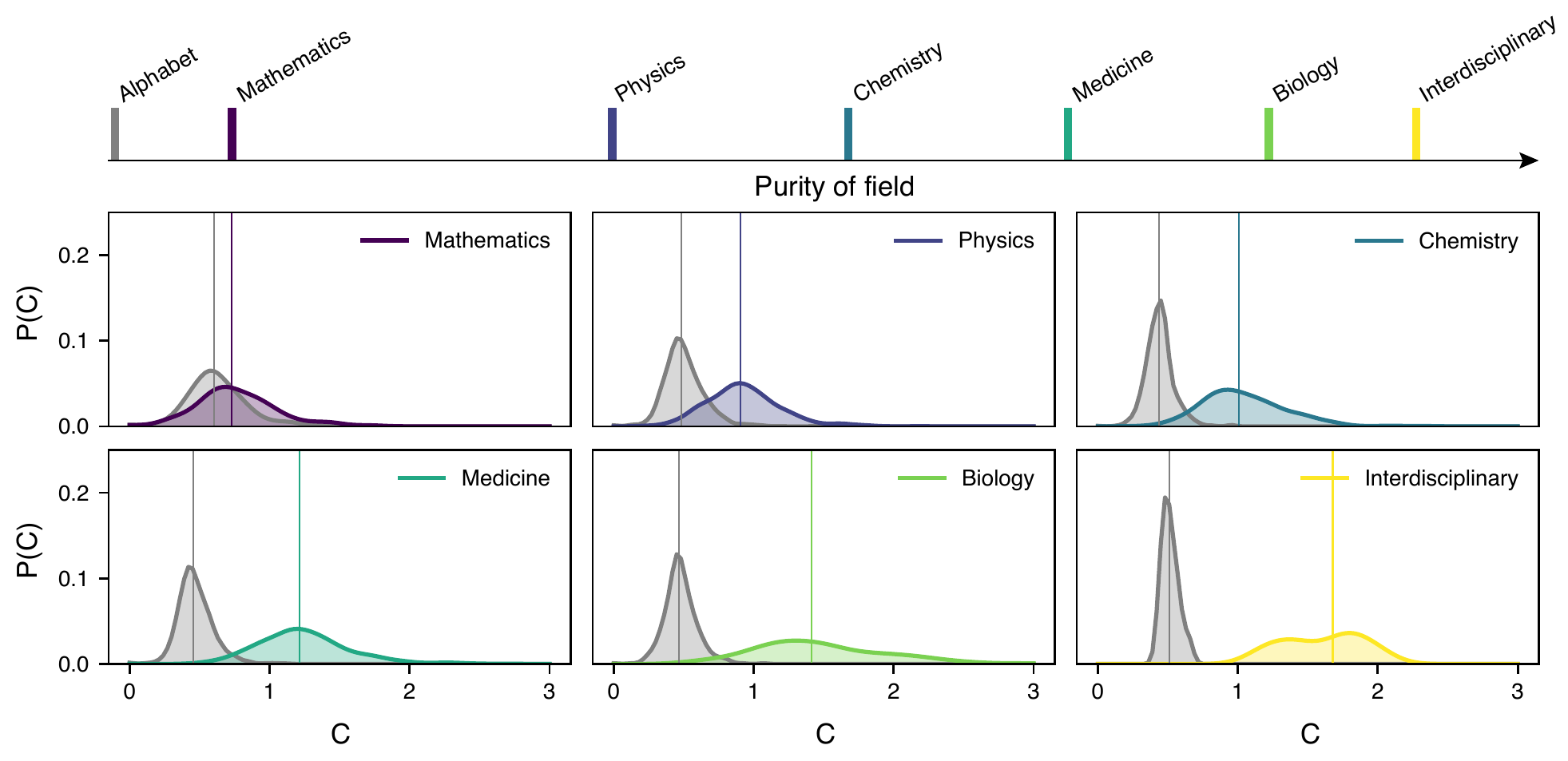}
\caption{\textbf{Comparison of chaperone effect between scientific fields.} Yearly distributions for the past 12 years are collapsed into single distributions and enable us to compare scientific fields (see SI Appendix section 2 and Fig. S1). For the different disciplines we find on average that: $\langle c/c_{\text{random}}\rangle_{\text{math}}\simeq 0.73$, $\langle c/c_{\text{random}}\rangle_{\text{physics}}\simeq 0.91$, $\langle c/c_{\text{random}}\rangle_{\text{chemistry}}\simeq 1.01$, $\langle c/c_{\text{random}}\rangle_{\text{medicine}}\simeq 1.21$, and $\langle c/c_{\text{random}}\rangle_{\text{biology}}\simeq 1.41$, while the effect for interdisciplinary journals is $\langle c/c_{\text{random}}\rangle_{\text{interdisciplinary}}\simeq 1.68$.
A Wilcoxon rank sum test, moreover, illustrates that the distributions are distinguishable $p\ll0.05$. The $C_{alphabet}$ distributions all peak around 0.5 because of the analytical properties of the null model (see SI Appendix section 1. for a proof). $C$ is represented by the colored distributions while $C_{\text{alphabet}}$ distributions are indicated in gray.}
\label{fig:chaperone_fields}
\end{figure*}

The proportion of these three kinds of author is substantially different depending on the journal. \emph{New England Journal of Medicine} (NEJM) is an example of a journal where the highest fraction of senior authors is new (Fig~1b, red line), signaling that repeat authorship is less common, i.e.~the medical community tends to submit only their most groundbreaking work to this high-impact general interest journal. 
As a point of contrast, we show \emph{Physical Review D} as an example of a journal where established PIs are predominant, a tendency which increases over in time (Fig~1b, blue line). 
This picture arises when some authors specialize in writing for a particular disciplinary journal, leading to a large fraction of repeated names in the PI spot.
In the bottom panel we show \emph{Nature}, a journal with an interdisciplinary audience, which has undergone a strong change over the past 10 years, with the fraction of \emph{new} authors dropping significantly. 
This indicates that it is becoming increasingly rare to publish as the senior author in \textit{Nature} without previous publishing experience in the journal.
A possible explanation for this development is an increasing number of authors specializing in writing papers for high-impact general audience journals, eschewing the more traditional pattern of publishing primarily in specialized journals and sending only selected results to high-impact multidisciplinary journals.

To understand the role of journal-specific experience, we investigate the \emph{chaperoned} authors more closely. 
\emph{Chaperoned} authors are senior authors, who have published in the journal previously as non-last authors (see Fig~1a).
Due to prior experience with the process, \emph{chaperoned} PIs already have gone through the intensive process of preparing a manuscript for a high-impact journal and absorbed tacit knowledge on how to frame the message appropriately for the journal audience, strike the right tone in the cover letter, structure the supporting information, the subtleties of how to constructively interact with editors, mastering layers of information that is usually invisible to those reading a paper.
Hence, the senior author acts as a chaperone simply through guiding the submission process. 
Having experienced the entire publication process once increases the chances of publishing in similar journals again, since the author is familiar with their particular idiosyncrasies.
In Fig~1b, the \emph{chaperoned} fraction hovers at around $0.1-0.2$ for all three journals, but showing an increasing trend over time for \textit{NEJM} and \textit{Nature}.
Thus in both \textit{NEJM} and \textit{Nature}, high impact journals with a wide audience, the fraction of \emph{new} authors decreases over time, while the fraction of \emph{chaperoned} authors slightly increases. In other words, it is becoming harder to publish in \emph{Nature} without having published in \emph{Nature} before.

We quantify the \emph{chaperone effect} $c$ for a journal by comparing the number of authors that over time have made the transition from a \textit{non-last} position to the \textit{last} position in the author order, with the number of last-authors that within the journal have never made such a transition over time. In other words, we compare the proportion between \textit{new} and \textit{chaperoned} PIs.
The chaperone effect captures the difficulty associated with publishing in a journal without previous experience with that journal. 
A chaperone effect of $c = 1$ implies that there is a balance between \emph{new} and \emph{chaperoned} authors. 
If the chaperone effect for a journal is greater than one ($c > 1$), it means that the fraction of \emph{new} authors is smaller than the fraction of \emph{chaperoned} authors and that, in order to publish in the journal in question, it is important to have a senior author act as a chaperone.
Conversely, if a journal has a chaperone effect smaller than one ($c < 1$), publication is easier for new authors. The specific value of $c$, however, is affected by field specific characteristics and publishing conventions, like typical team size and individual productivity \cite{way2017misleading}. It also does not take into account the fact the an author can make the transition from non-last to last position randomly. For these reasons, $c$ of different fields cannot be directly compared.

To correct for these caveats in the quantification of the chaperone phenomenon through $c$ and to be able compare the importance of apprenticeship across the sciences, we compare the observed values of \textit{new} and \textit{chaperoned}, with those occurring in two null models \cite{radicchi2008universality, radicchi2012testing}.
Firstly, we consider a system where the ordering of author names is not relevant~\cite{sinatra2016quantifying}.
Therefore, we compare $c$ to $c_{\text{random}}$, where $c_{\text{random}}$ is the ratio obtained in a null model where we have randomly permuted the order of author names in each paper.
We call $C = c/c_{\text{random}}$ the \emph{magnitude} of the chaperone effect. 
Note that the magnitude $C$ cannot be affected by team-size and individual volume productivity, as these are preserved in the randomization. 
However, $C$ does capture significant changes in the order of authors in respect to the random ordering. 
In general, the chaperone phenomenon occurs when $C > 1$, \emph{i.e.}~when the transition \textit{non-last}$\rightarrow$\textit{last} is more frequent than the appearance of new authors in the last position in a statistically significant way. 
Secondly, in some fields (e.g. mathematics), alphabetical author-sorting is an important convention. 
Therefore, we also compare $c$ to $c_{\text{alphabet}}$, which is based on a system where all author-lists are sorted alphabetically~\cite{mathstatement}. 
Based on this second model, we construct $C_{\text{alphabet}} = c/c_{\text{alphabet}}$. 
Values of $C_{\text{alphabet}}$ are typically smaller than one (see SI Appendix section 1.). In a nutshell, the deviation of $c$ from $c_{\text{random}}$ and $c_{\text{alphabet}}$ provides the magnitude of the chaperone effect, stripped of any confounding effects (see Methods).

In Fig.~2, we show the distribution of $C$ as well as $C_{\text{alphabet}}$ for the five fields mentioned above and for interdisciplinary journals.
This figure is in line with the collective intuition about `the purity of sciences'~\cite{xkcd}.
Mathematics shows very few signs that experience influences the transition between junior and senior levels.
This is likely in part due to the fact that authorship conventions in mathematics dictate alphabetical order for all publications~\cite{mathstatement}. 
We see the magnitude of the chaperone effect growing across physics, chemistry, and medicine, with the strongest effect within biology and general-topic journals. 
For these fields, there is a clear relationship between having published in them as a junior researcher and the probability of publishing in them as PI, illustrating that experience with publication is important for transitioning between junior and senior authorship within high-impact journals.

Assessing the existence of an unbalanced proportion of chaperoned and new last-authors prompts an important question: How does publishing in a journal as a non-last author impact your odds for one day publishing as last author?
Since we do not have access to statistics for rejected papers, we are unable to answer that question exactly. 
We can, however, answer a closely related question, namely: How does the probability of transitioning to last author change as a function of number of occurrences as a non-last author? 
In Fig.~3a we see that, in the case of \emph{Nature}, this probability grows significantly from $10\%$ after one publication to nearly $20\%$ after four publications as non-last author (for the study of chaperone effect in PNAS, see SI Appendix figs.~S3 and S4).
In contrast, the same transition probability is $25\%$ in the case of the highly disciplinary \emph{Physical Review D}, and does not change with additional publications as non-last author.

There is a critical aspect of the chaperone effect that we have not yet explored: does experience with publication within a certain journal play a role in the scientific impact of subsequent papers as PI published in the same venue? 
Could it be that \textit{new}, \textit{chaperoned} and \textit{established} last-author papers receive different levels recognition from the scientific community?
Our initial hypothesis was that papers authored by \textit{new} PIs might have higher impact, since their lower odds of being published might signal a higher significance of the reported discoveries for the scientific community. 
To test this hypothesis, we quantified the impact of each paper by measuring $c_5$, its citations after 5 years from publication, a measure that is not affected by the specific field citation dynamics~\cite{wang2013quantifying}. 
This allowed us to directly calculate the average impact over time for three categories of papers, those with chaperoned, established and new PIs. 
In the case of \emph{Nature} the result is striking (Fig.~3b). 
We find that papers with established and chaperoned PIs have indistinguishable impact. 
Contrary to what we expected, however, papers authored by \emph{new} PIs in \emph{Nature} receive on average only half the citations of papers authored by \textit{chaperoned} and \textit{established} PIs, indicating a systematically lower scientific impact. 
The same pattern is observed in the entire group of interdisciplinary journals, suggesting this pattern is consistent in these venues with  high selection pressure and only a small fraction of all scientists manage to publish as PI.
In more specialized field-specific journals, a difference can be also present, but the differences between the three categories of authors tend to be smaller (see SI Appendix).
Thus, our findings suggests that experience of publishing within specific journals can play an important role in acquiring long-term scientific impact.
\begin{figure}
	\centering
	\includegraphics[width=1\linewidth]{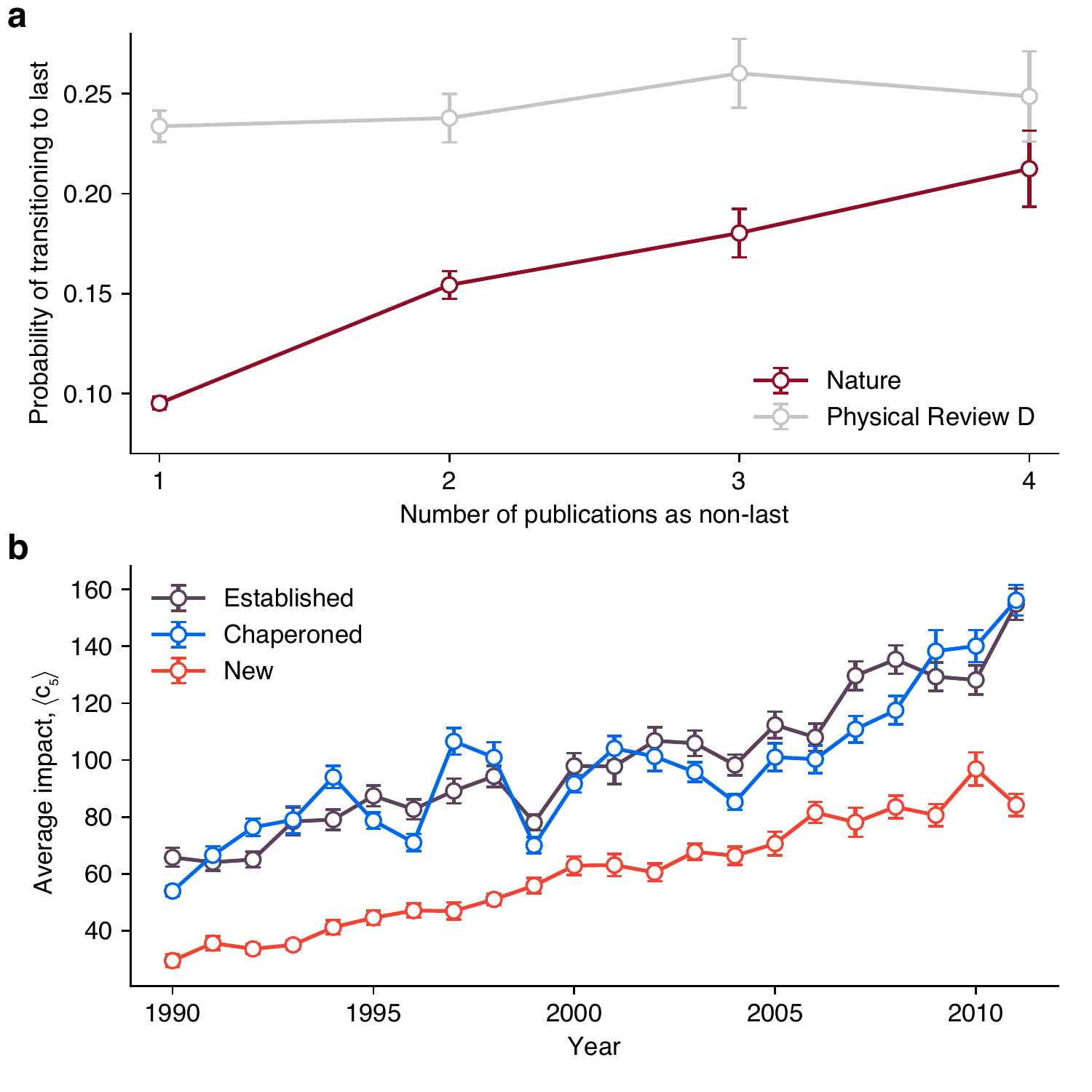}
	\caption{\textbf{The advantages of chaperoned and established PIs}. \textbf{a,} The probability of transitioning to last author as a function of number of occurrences as non-last author for a specialized journal, \textit{Physical Review D} and an interdisciplinary journal, \textit{Nature}. \textbf{b,} Average impact of papers in \textit{Nature}, quantified with citations after 5 years from publication ($c_5$), for papers authored by \emph{new}, \emph{chaperoned} and \emph{established} last authors.}
\end{figure}

Taken together, our results add new piece to the puzzle of how mentor-protegee relations function more generally~\cite{chao1992formal, scandura1992mentorship, higgins2001reconceptualizing, allen1997mentor, singh2009gets, malmgren2010role}, where a full picture of the relation will also draw on understanding of how teams are assembled and produce knowledge~\cite{guimera2005team, wuchty2007increasing}.  
In this sense, additional research is needed to understand the complex processes that drive the differences between \emph{new} and \emph{chaperoned authors}.
By focusing on the role of experience within journals and fields, we deliberately average over authors with very different levels of success and do not account for the fact that good proteges tend to find good mentors; nor do we include the fact many young authors leave science altogether. 
Therefore, it is important to stress that our results are not designed to answer the deeper questions on mentor-protege rules, but point to the general structures in how knowledge needed to write for certain journals is different across the sciences, with high-impact, interdisciplinary journals showing a particularly strong effect.

Thus, while the available data here do not allow us to strictly pinpoint which facet of experience is most important in order to succeed in science, or which share of a senior authors's apprentices are successfully chaperoned \cite{clauset2017data}, we have demonstrated that the chaperone effect does indeed exist, showing that the ability to publish in certain venues is something that junior scientists learn from senior colleagues. 
Further, we have demonstrated that apprenticeship is not just about membership in the exclusive club of having-published-in-\emph{Nature} or another prestigious journal, but papers by \emph{chaperoned} authors have greater scientific impact than papers by \emph{new} PIs.
In addition, we show that the magnitude of the chaperone effect varies across scientific fields~\cite{xkcd}. 
The chaperone effect is most strongly expressed in prestigious multidisciplinary journals, demonstrating that the highly specialized skill-set required in order to publish in these venues is passed along more strongly than any field-specific expertise. 

\matmethods{
\subsection*{Data}
We use publication data provided by the \textit{Web of Science} database (www.webofknowledge.com), purchased for research purposes by some of the authors of this publication in 2013. The database includes several types of scientific outputs such as articles, letters, reviews, editorials and abstracts from 1898 to 2012 across more than 22,000 scientific journals from broad domains, resulting in a set of more than 50 millions papers. For each paper, the dataset includes more information on the date of publication (month, day, year), the journal name and journal issue, author names with the order they appear on the article, their affiliations, and the references towards past articles indexed in the database. For \textit{Nature} we have downloaded the full publication history using the \textit{Nature} opensearch API. 

For our analysis, we focused on publications from 1960 to 2012 published in interdisciplinary journals (\textit{Nature}, \textit{Science}, and \textit{PNAS}), as well as in journals associated to five distinct scientific fields: \textit{Medicine}, \textit{Biology}, \textit{Mathematics}, \textit{Chemistry} and \textit{Physics}. 
To identify the journals belonging to each category, we first parsed dedicated Wikipedia pages containing lists of journal names associated to specific scientific fields and then matched these with the journals in the database \cite{sinatra2015century}. 
In total we identified 97 \textit{biology},  337 \textit{medicine}, 243 \textit{physics}, 248 \textit{mathematics}, 138 \textit{chemistry}, and 3 \textit{interdisciplinary} journals.

Next, we extracted the publications associated to each of these categorized journals.
To ensure to deal with original research, we collected only publications labeled as \textit{Articles}, \textit{Letters} and \textit{Reviews} and that did not have a title containing the terms \textit{comment}, \textit{reply}, \textit{errata}, or \textit{retracted article}. Moreover, in order to have enough statistics, only the categorized journals fulfilling the following criteria were taken into account for our analysis:
\begin{itemize}
\item The collected publications associated to the journal are spanning a period of at least 10 years;
\item At least 1,000 collected publications were published in the journal overall;
\item At least 100 collected publications were published each year in the journal.
\end{itemize}
After this preprocessing, our data amounts to \textit{(i)} 795,558 publications from 40 journals in \textit{biology}, \textit{(ii)} 1,350,936 publications from 128 journals in \textit{medicine}, \textit{(iii)} 1,753,641 publications from 117 journals in \textit{physics}, \textit{(iv)} 208,223 publications from 26 journals in \textit{mathematics}, \textit{(v)} 1,341,150 publications from 72 journals in \textit{chemistry} and \textit{(vi)} 251,294 publications from \textit{Nature}, \textit{Science} and \textit{PNAS}.
Data about the proportion of \emph{new}, \emph{established} and \emph{chaperoned} PIs over time, and the values of $c$, $C$ and $C_{alphabet}$ are provided for each journal on GitHub (https://github.com/SocialComplexityLab/chaperone-open). Raw data from Web of Science cannot be shared publicly on the web, but we offer the possibility to reproduce our results starting from raw records by spending a research visit at Northeastern University or Central European University where the data is accessible. Data about the journal \textit{Nature} can be downloaded for free from \textit{Nature} opensearch (https://www.nature.com/opensearch/).

\subsection*{Author name disambiguation}
We formatted all author names present on the collected publications to lower case and converted their names into their first letter only. An author named "\textit{John~Smith}" or "\textit{Mary Suzy Johnson}" would thus be converted to the format "\textit{smith,j}" or "\textit{johnson,ms}", respectively. 
We considered the sequence of publications within the same journal and authored by an identical formatted name to correspond to a same individual. We expect errors induced by homonyms, i.e. distinct individuals that share the same formatted name, to be low as we only compare names within the same journal. An error can thus only occur if two distinct individuals share the same formatted name and evolve in the same scientific field, i.e. the same journal, which is already an accurate disambiguating feature \cite{torvik2009author}.

\subsection*{Robustness of results to alphabetic ordering}
In certain scientific fields it is common to order authors alphabetically \cite{waltman2012empirical}.
As such, to understand how this affects the results, we perform two versions of our analysis: one, taking all publications into account, and two a version where we have disregarded publications where authors are alphabetically ordered.
This removes $17.7\%$ of all publications within \textit{Biology}, $14.4\%$ within \textit{Medicine}, $30.9\%$ within \textit{Physics}, $75.1\%$ within  \textit{Mathematics}, $23.3\%$ within  \textit{Chemistry}, and $20.8\%$ within \textit{interdisciplinary} journals.
Note that these numbers include publications where the authors are ordered by choice, but also publications where this occurred by chance.
Nonetheless, our conclusions are robust for both data sets, consistently with the result shown in Fig.~\ref{fig:chaperone_fields} that there is a significant difference between observed $C$ and that of the alphabetical null model $C_{alphabet}$ (see SI Appendix fig.~S5).
}

\showmatmethods 

\acknow{This work was supported by Air Force Office of Scientific Research grants FA9550-15-1-0077 and FA9550-15-1-0364 (A.-L.B. and R.S.), The European Commission, H2020 Framework program, Grant 641191 CIMPLEX, The Templeton Foundation (R.S., A.-L.B.), and the ITI project `Just Data' funded by Central European University (R.S.), The Villum Foundation (S.L.), The Independent Research Fund Denmark (S.L.). R.S. thanks Michael Szell for useful discussions and feedback, and Alex Gates for support with the Web of Science data.
}

\showacknow 


\bibliography{biblio}

\end{document}